\documentclass[fleqn,11pt]{article}

\usepackage{amsmath}
\usepackage{amssymb}

\usepackage{amsfonts,amssymb,cite}
\usepackage{epsf,graphicx}

\def\noi{\noindent}
\def\jnumber#1#2{\thispagestyle{empty} \noi\unitlength=1mm
    	\begin{picture}(178,10)
            \put(177,15){\llap{\large\it Grav. Cosmol. No.\,#1, #2}}
                    \end{picture}}

\newcommand{\Title}[1]{\noi {{\Large\bf #1}}\\[1ex]}

\newcommand{\Author}[2]{\noi{\bf #1}\\[2ex]\noi{\normalsize\it #2}\\}

\newcommand{\Rec}[1]{\noi {\it Received #1} \\}

\newcommand{\Abstract}[1]{\vskip 2mm \begin{center}
        \parbox{16.4cm}{\small\noi #1} \end{center}\medskip}

\newcommand{\foom}[1]{\protect\footnotemark[#1]}
\newcommand{\foox}[2]{\footnotetext[#1]{#2}\addtocounter{footnote}{1}}
\def\email#1#2{\footnotetext[#1]{e-mail: #2}\addtocounter{footnote}{1}}

\def\Talk{\foox 1 {Talk given at the International Conference RUDN-10,
	   June 28 --- July 3, 2010, PFUR, Moscow}}

\def\nqq{\hspace*{-2em}}

\def\cm{\hspace*{1cm}}
\def\inch{\hspace*{1in}}

\def\Jl#1#2{#1 {\bf #2},\ }

\def\ApJ#1 {\Jl{Astroph. J.}{#1}}
\def\CQG#1 {\Jl{Class. Quantum Grav.}{#1}}
\def\DAN#1 {\Jl{Dokl. AN SSSR}{#1}}
\def\GC#1 {\Jl{Grav. Cosmol.}{#1}}
\def\GRG#1 {\Jl{Gen. Rel. Grav.}{#1}}
\def\JETF#1 {\Jl{Zh. Eksp. Teor. Fiz.}{#1}}
\def\JETP#1 {\Jl{Sov. Phys. JETP}{#1}}
\def\JHEP#1 {\Jl{JHEP}{#1}}
\def\JMP#1 {\Jl{J. Math. Phys.}{#1}}
\def\NPB#1 {\Jl{Nucl. Phys. B}{#1}}
\def\NP#1 {\Jl{Nucl. Phys.}{#1}}
\def\PLA#1 {\Jl{Phys. Lett. A}{#1}}
\def\PLB#1 {\Jl{Phys. Lett. B}{#1}}
\def\PRD#1 {\Jl{Phys. Rev. D}{#1}}
\def\PRL#1 {\Jl{Phys. Rev. Lett.}{#1}}

\def\lal{&&\nqq {}}

\def\beq{\begin{equation}}
\def\eeq{\end{equation}}
\def\bear{\begin{eqnarray}}
\def\bearr{\begin{eqnarray} \lal}
\def\ear{\end{eqnarray}}
\def\earn{\nonumber \end{eqnarray}}

\def\nnn{\nonumber\\ \lal }

\def\d{\partial}

\def\eps{\varepsilon}

\begin{document}
\twocolumn[
\jnumber{2}{2011}

\Title{Gravitation as an effect of nonlinear electrodynamics\foom 1}

\Author{Alexander A. Chernitskii\foom 2}
   {Friedmann Laboratory of Theoretical Physics, St. Petersburg, Russia;\\
    State University of Engineering and Economics,
        ul. Marata 27, St. Petersburg 191002, Russia}

\Rec {October 4, 2010}

\Abstract
  {We consider an approach to unification of the gravitational and
   electromagnetic interactions based on the existence of an effective
   Riemannian space in nonlinear electrodynamics. In the context of this
   approach, we discuss the possibility of screening the gravitational field.
   }


\bigskip
] 
\Talk
\email 2 {AAChernitskii@mail.ru,\\ AAChernitskii@engec.ru}

\section{Introduction}

  The author's works [1, 2] have suggested an approach to solving the
  problem of unification of gravitation and electromagnetism on the basis
  of the existence of an effective Riemannian space for interacting solitons
  in nonlinear electrodynamics. In [1], Mie's model of nonlinear
  electrodynamics was considered, and in [2] that of Born-Infeld.
  A characteristic feature of the present approach to the unification
  problem is that gravitation is secondary with respect to electromagnetism,
  or one can say that it is induced by nonlinear electromagnetism.

  Different aspects of this approach have been discussed in [3--7] and
  elsewhere.

  Under such an approach, it is natural to expect that some aspects of
  electromagnetism should manifest themselves in gravitation as well. In
  particular, one can raise the question of screening the gravitational
  field similar to the well-known electromagnetic field screening.

\section{The field model and electromagnetic particles}

  As the field of the model, we take the 4-vector field $A_{\mu}$ (Greek
  indices take the values $0,1,2,3$), since it is this vector with respect
  to which the action functional is varied in electrodynamics. The
  electromagnetic tensor is defined in the usual way,
\beq                                               \label{ch-56419188}
    F_{\mu\nu} = \frac{\d A_\nu}{\d x^\mu} - \frac{\d A_\mu}{\d x^\nu}.
\eeq

  As the variational principle of the model, the following generally
  covariant variational principle is taken:
\bearr
    \delta\int\!\!\sqrt{|\det(G_{\mu\nu})|}\;(\mathrm{d}x)^{4} = 0,
\nnn            \label{ch-57803273}
    \inch       G_{\mu\nu} = g_{\mu\nu} + \chi^2\,F_{\mu\nu}.
\ear
  where $g_{\mu\nu}$ is the metric tensor of flat space-time,
  $(\mathrm{d}x)^4\equiv  \mathrm{d}x^0\mathrm{d}x^1\mathrm{d}
  x^2\mathrm{d}x^3$, and $\chi$ is a dimensionful constant.

  This variational principle was considered in a somewhat more general form
  by A.S. Eddington [8] and A. Einstein [9]. M. Born and L. Infeld [10]
  have considered an electrodynamic model corresponding to the variational
  principle (2).

  The set of equations obtained from the variational principle (2) has the
  form
\beq                    \label{ch-35357121}
    \frac{\d}{\d x^\mu}\,\sqrt{|g|}\; f^{\mu\nu} {}={} 0,
\eeq
  where
\bearr                    \label{ch-65713937}
    f^{\mu\nu} {}\equiv {} \frac{\chi^{-2}\,\d{\cal L}}{\d(\d_\mu A_\nu)}
    = \frac{1}{\cal L}\,\left(F^{\mu\nu} -
    \frac{\chi^2}{2}\,\cal J\,\eps^{\mu\nu\sigma\rho}\,
    F_{\sigma\rho}\right),
\nnn
\\ \lal
     {\cal L}\equiv \sqrt{|\,1 - \chi^2\,{\cal I}- \chi^4\,{\cal J}^2\,|},
\nnn
    {\cal I} \equiv F_{\mu\nu}\,F^{\nu\mu}/2,
\nnn
    {\cal J} \equiv \eps_{\mu\nu\sigma\rho}\, F^{\mu\nu} F^{\sigma\rho}/8,
\nnn
    \eps_{0123} \equiv \sqrt{|g|}, \cm \eps^{0123}  = - 1/\sqrt{|g|}.
\earn
  The symmetric stress-energy tensor of the model has the form
\beq                                                \label{ch-71416389}
    T^\mu_{.\nu} \equiv  \left(f^{\mu\rho}\,F_{\nu\rho}  {}-{}
    \chi^{-2}\,\left({\cal L} {}-{} 1\right)\,\delta^\mu_\nu\right)/4\pi.
\eeq

  The characteristics equation in this model has the following remarkable
  form:
\beq                                    \label{ch-CharEq}
     \tilde{g}^{\mu\nu}\,\frac{\d\Phi}{\d x^\mu}\,\frac{\d \Phi}{\d x^\nu}=0,
\eeq
  where $\Phi (x^\mu)=0$ is the equation of the characteristic surface,
\beq                                            \label{ch-73144072}
    \tilde{g}^{\mu\nu} \equiv g^{\mu\nu} - 4\pi\,\chi^2\,T^{\mu\nu}.
\eeq
  Here $T^{\mu\nu}$ is the stress-energy tensor of the
  form (\ref{ch-71416389}).

  The set of equations in this model has spatially localized solutions
  which are attributed to electromagnetic particles. The simplest solution
  of this kind, possessing spherical symmetry, has been considered by Born
  and Infeld as a model of the electron [9]. This solution has a finite
  total energy (mass). However, since the spherically symmetric solution
  does not possess a magnetic moment and a total angular momentum (spin),
  it cannot be related to the real electron. More complex solutions with
  axial symmetry can have a mass, a spin and a magnetic moment [2, 12].

  In addition to a static part, particlelike solutions should have a rapidly
  oscillating part responsible for the wave properties of particle and, as
  it turns out in the present approach, for the gravitational interaction
  [13].

\section{The effective Riemannian space and gravitation}

  In an investigation of a multiparticle solution using a perturbation
  method, one takes as an initial approximation a sum of one-particle
  solutions with trajectories determined from the condition of total
  momentum conservation in the localization domains of solitonic particles
  [2]. n this way one obtains the electromagnetic interaction of solitonic
  particles. This interaction originates from the static part of the
  solitonic solution (in its own reference frame).

  To explain the gravitational interaction in this approach, it is necessary
  to take into account the rapidly oscillating part of the soliton under
  study and the field of remote solitons. In addition, it is necessary to
  take into account the rapidly oscillating electromagnetic background.

  In the initial approximation, the light wave and the rapidly oscillating
  part of a solitonic particle propagate along geodesics of the effective
  Riemannian space with the metric $\tilde{g}^{\mu\nu}$, depending on the
  field of remote solitons. This follows from the characteristic equation
  (\ref{ch-CharEq}) [2].

  In the initial approximation, the static part and the rapidly oscillating
  part of a massive solitonic particle are connected. In this way, the
  static part is as though dragged by the rapidly oscillating part.

  For a correct behavior of the effective metric (i.e., in order to obtain
  the Newtonian potential), the radiation background must be taken into
  consideration [3, 13].

  Thus the gravitational interaction is stipulated by the rapidly
  oscillating component of solitonic solutions.

\section{On possible screening\\ of gravity}

  Since in the framework of the present approach the gravitational
  interaction is a manifestation of the electromagnetic field nonlinearity,
  it is logical to consider the opportunity of screening the gravitational
  field similar to electromagnetism. Herewith, since gravity is related to
  the rapidly oscillating part of the solitonic solutions, one should
  consider a shield for electromagnetic waves as a prototype gravitational
  screen.

  As shown in [13], the Newtonian potential in the effective metric created
  by a certain soliton can emerge due to an interaction between the rapidly
  oscillating part of the soliton with the radiation background. If one
  places the soliton into a perfectly reflecting sphere, there will be no
  interaction between the soliton and the radiation background. It should be
  noted that outside the sphere the radiation background will excite the
  corresponding spherical modes of standing electromagnetic waves.  Their
  interaction with the background can again give a Newtonian potential in
  the effective metric. However, if we suppose that the radiation background
  has a sufficiently small magnitude, then the gravitational field outside
  the sphere will be weaker than if the spherical screen were absent.

  Moreover, if one places a plane plate of finite size closely to a soliton,
  then, in the shade region for the solitonic radiation (i.e., for the
  rapidly oscillating part of the solution), one can also expect weakening
  of the soliton's gravitational field.

  Of interest is certainly the question of the existence of substances from
  which one could fabricate such a reflecting screen. To answer it, one
  should above all determine the order of magnitude of the frequency of the
  rapidly oscillating parts of solitonic solutions corresponding to real
  particles. This frequency is determined from the well-known relations of
  quantum physics:
\beq                    \label{ch-31021689}
    mc^2 = h\,\nu\quad \Longrightarrow\quad
            \nu = \frac{m\,c^2}{h} = \frac{c}{\lambda_C},
\eeq
  where $m$ is the particle mass, $c$ is the speed of light in vacuum, $h$
  is the Planck constant, $\nu$ is the frequency, and $\lambda_C$ is the
  Compton wavelength of the particle.

  Bearing in mind the Compton wavelengths of the electron
  ($\approx 2.4\cdot 10^{-3}$ nm) and proton ($\approx 1.3\cdot 10^{-6}$
  nm), let us estimate the frequency:
\beq                                 \label{ch-36351183}
    \nu \sim 10^{20} \div 10^{23} \quad (\text{Hz}).
\eeq
  Such frequencies and the corresponding electromagnetic wavelengths are
  characteristic of gamma radiation ($\lambda < 5\cdot 10^{-3}$ nm).

  Thus {\it to obtain a gravitational screen in the present approach it
  is necessary to have a mirror for gamma rays}.

  Gamma rays are known to possess a high penetrating power in usual
  substances, and creating such a mirror can now seem to be an unreal idea.
  One should not, however, exclude the opportunity of obtaining, say, a
  nano-material that would possess the necessary properties.

\section{Conclusion}

  In the framework of the theory under consideration, there is in principle
  an opportunity of a gravitational screening effect. Such screening may be
  realized with the aid of materials with unusual properties.

  From this viewpoint, it seems to be expedient to carry out the
  corresponding experimental tests of newly obtained (nano-)materials for the
  necessary properties (from the viewpoint of the present approach) for
  observing the gravitational screening effect.


\small
\begin{thebibliography}{10}

\bibitem{ch-Chernitskii1992}  
    A.~A. Chernitskii.
    {\it Long-range interaction of four-vector field solitons of the
    Minkowskian space,} Theoret. Math. Phys. {\bf 90} (3), 260 (380
    in Russian) (1992).

\bibitem{ch-Chernitskii1999}  
    A.~A. Chernitskii, {\it Dyons and interactions in nonlinear
    (Born-Infeld) electrodynamics},
    J. High Energy Phys. {\bf 12}, Paper 10 (1999); hep-th/9911093.

\bibitem{ch-Chernitskii2002b}  
    A.~A. Chernitskii, {\it Induced gravitation as nonlinear
    electrodynamics effect}, Grav. Cosmol. {\bf 8},
        Supplement, 157--160 (2002); gr-qc/0211034.

\bibitem{ch-Chernitskii2004a}  
    A.~A. Chernitskii, {\it Born-Infeld equations.} In:
    {\em Encyclopedia of Nonlinear Science}, ed. A.~Scott,
    (Routledge, New York --- London, 2004), 67--69; hep-th/0509087.

\bibitem{ch-Chernitskii2006c}  
    A.~A. Chernitskii, {\it Gravitation and electromagnetism in a
    theory of a unified four-vector field},
    \GC {12} 130--132 (2006);  hep-th/0609204.

\bibitem{ch-Chernitskii2008a} 
    A.~A. Chernitskii,
    {\it Gravitation as a vacuum nonlinear electrodynamics effect.}
    In: {\em On Recent Developments in Theoretical and Experimental
    General Relativity, Gravitation and Relativistic Field Theories}
    (eds. R.~H.Kleinert and R. T. Jantzen, World Scientific, 2008),
    1236--1238; arXiv: 1006.5682.

\bibitem{ch-Chernitskii2009a} 
    A.~A. Chernitskii,
    {\it On unification of gravitation and electromagnetism in the
    framework of a general-relativistic approach},
    \GC {15} 151--153 (2009); arXiv: 0907.2114.

\bibitem{ch-Eddington1924} 
    A.~Eddington, {\em The Mathematical Theory of Relativity}
    (Cambridge, 1924).

\bibitem{ch-Einstein1923dE} 
    A.~Einstein, {\it Zur allgemeinen Relativit\"atstheorie},
    Sitzungsber. preuss. Akad. Wiss., phys.-math., 32--38. Kl., 1923.

\bibitem{ch-BornInfeld1934a}  
    M.~Born and L.~Infeld.
    {\it Foundation of the new field theory},
    Proc. Roy. Soc. A {\bf 144}, 425--451 (1934).

\bibitem{ch-Chernitskii1998b}  
    A.~A. Chernitskii,
    {\it Light beams distortion in nonlinear electrodynamics},
    J. High Energy Phys. {\bf 11}, Paper 15 (1998); hep-th/9809175.

\bibitem{ch-Chernitskii2006a}  
    A.~A. Chernitskii,
    {\it Mass, spin, charge, and magnetic moment for electromagnetic
    particle}.  In: {\em XI Advanced Research Workshop on High Energy
    Spin Physics (DUBNA-SPIN-05) Proceedings}, (eds. A.~V. Efremov and
    S.~V. Goloskokov, JINR, Dubna, 2006), 234--239; hep-th/0603040.

\bibitem{ch-Chernitskii2006b}  
    A.~A. Chernitskii,
    {\it Linear waves around static dyon solution of nonlinear
    (Born-Infeld) electrodynamics},
    Hadronic Journal {\bf 29}, 497--528 (2006); hep-th/0602079.

\end{thebibliography}
\end{document}